\documentclass[prb,longbibliography,aps,reprint,showpacs,raggedbottom,nofootinbib,superscriptaddress,floatfix]{revtex4-2}

\usepackage{graphicx}
\usepackage{dcolumn}
\usepackage{bm}
\usepackage{hyperref}
\usepackage{gensymb}
\usepackage{siunitx}
\usepackage{amsmath}
\usepackage{amsfonts}
\usepackage{xcolor}
\usepackage{array,multirow}
\usepackage[normalem]{ulem}
\usepackage{textcomp}
\newcommand{\ignore}[1]{}

\begin{document}

\title{Emission Dynamics of Rydberg Excitons in $\mathbf{\mathrm{Cu_2O}}$: Distinguishing Second Harmonic Generation from Secondary Emission}

\author{Kerwan Morin}
\thanks{These authors contributed equally to this work.}
\affiliation{Univ Toulouse, INSA Toulouse, CNRS, LPCNO, Toulouse, France}
\author{Poulab Chakrabarti}
\thanks{These authors contributed equally to this work.}
\affiliation{Univ Toulouse, INSA Toulouse, CNRS, LPCNO, Toulouse, France}
\author{Delphine Lagarde}
\affiliation{Univ Toulouse, INSA Toulouse, CNRS, LPCNO, Toulouse, France}
\author{Maxime Mauguet}
\affiliation{Univ Toulouse, INSA Toulouse, CNRS, LPCNO, Toulouse, France}
\author{Sylwia Zielińska - Raczyńska}
\affiliation{Department of Physics, Bydgoszcz University of Science and Technology, Bydgoszcz, Poland}
\author{David Ziemkiewicz}
\affiliation{Department of Physics, Bydgoszcz University of Science and Technology, Bydgoszcz, Poland}
\author{Xavier Marie}
\affiliation{Univ Toulouse, INSA Toulouse, CNRS, LPCNO, Toulouse, France}
\author{Thomas Boulier}
\email{boulier@insa-toulouse.fr}
\affiliation{Univ Toulouse, INSA Toulouse, CNRS, LPCNO, Toulouse, France}

\date{\today}

\begin{abstract}
Rydberg excitons in $\mathrm{Cu_2O}$ simultaneously give rise to two very different optical responses under resonant two-photon excitation: a coherent second-harmonic signal mediated by the excitonic second order susceptibility tensor $\chi^{(2)}$, and a secondary emission originating from the radiative decay of real exciton populations. Distinguishing these two channels is essential for interpreting nonlinear and quantum-optical experiments based on high-$n$ states, yet their temporal, spectral, and power-dependent signatures often overlap. Here we use time-resolved resonant two-photon excitation to cleanly separate SHG and SE and to map how each depends on $n$, temperature, excitation power, and crystal quality. This approach reveals the markedly different sensitivities of the two processes to phonons, defects, and many-body effects, and establishes practical criteria for identifying SE and SHG in a wide range of experimental conditions. Our results provide a unified framework for interpreting emission from Rydberg excitons and offer guidelines for future studies aiming to exploit their nonlinear response and long-range interactions.
\end{abstract}

\maketitle

\section{Introduction}

Rydberg excitons in Cu$_2$O offer a unique opportunity to explore optical nonlinearities in a regime where material excitations behave with near-atomic character while remaining embedded in a bulk crystal~\cite{assmann2020semiconductor}. With principal quantum numbers observed up to $n=30$~\cite{versteegh2021giant} and large Bohr radii on the order of micrometers, these giant excitons exhibit strong long-range interactions~\cite{kazimierczuk2014giant,walther2018interactions} and substantial optical nonlinearities~\cite{heckotter2021asymmetric,morin2022self}, motivating experiments in coherent many-body physics and quantum optics. For example, Rydberg excitons can be engineered into cavity polaritons possessing strong light-matter coupling and giant nonlinearities~\cite{walther2018giant,makhonin2024nonlinear} while microwave fields have been shown to strongly couple different Rydberg levels~\cite{gallagher2022microwave}, revealing coherent Kerr-like optical nonlinearities~\cite{pritchett2024giant}. These recent advances sparked strong interest in view of quantum technologies~\cite{assmann2020semiconductor,heckotter2024rydberg}. In parallel, high-resolution absorption, photoluminescence and nonlinear spectroscopy continue to reveal the exquisitely subtle exciton features present in this exemplar material~\cite{thewes2015observation,rogers2022high}. In the latter method, a two-photon excitation generates a signal at twice the pump energy, which is massively enhanced whenever the two-photon excitation is resonant with an exciton state. This principle has been exploited in so-called "SHG spectroscopy" to explore the dark (even parity) excitons in Cu$_2$O~\cite{Mund2018high,rogers2022high}. This is a powerful tool to understand Rydberg excitons and a crucial step in their exploitation for future quantum technologies. However, the fundamental relationship between the coherent nonlinear optical response of these high-$n$ states and their population dynamics remains largely unexplored. Understanding how these two facets evolve across the Rydberg series, and how they respond to local defects, thermal fluctuations and population density, is essential both for interpreting current experiments and for assessing the feasibility of Rydberg-based optical functionalities in solids. 

On the one hand, $\mathrm{Cu_2O}$ is centrosymmetric (point group $O_h$), so the bulk electric‐dipole $\chi^{(2)}$ vanishes. Nevertheless, second‐harmonic generation (SHG) is allowed through the higher-order electric‐quadrupole / magnetic‐dipole pathways along low-symmetry axis (like [111])~\cite{farenbruch2020prb}. It becomes resonantly enhanced at exciton energies~\cite{Mund2018high}, with well‐defined polarization selection rules. Therefore, the process we call SHG here probes a virtual, coherence‑based $\chi^{(2)}$ response that ends with the pump pulse. It is resonantly enhanced by excitons but leaves no long‑lived exciton population. On the other hand, two photons may also be absorbed to excite an even-parity exciton (e.g., a $S$ or $D$ state). This real exciton population then relaxes and decays radiatively. In Cu$_2$O, this “secondary” luminescence (which we refer to as \textit{secondary emission}, SE) typically occurs through electric-quadrupole transitions and its dynamics is controlled by the exciton lifetime~\cite{Chakrabarti2025,zielinska2025quantum}. Thus, under two-photon excitation, the detected light is a mixture of instantaneous SHG and delayed exciton luminescence. However, as both depend on the resonant exciton properties, they share many observables and are typically hard to separate.

Distinguishing SHG from secondary emission can be crucial for the correct interpretation of advanced experiments on Cu$_2$O Rydberg excitons. For example, recent work revealed coherent oscillations in the two-photon-induced emission~\cite{Chakrabarti2025}. Correctly interpreting this work requires to separate the two contributions and only focus on the SE, as only it carries the exciton late-time coherence information. More generally, the existence of two twin channels blurs the interpretation of the signal strength (and therefore quantities such as the oscillator strength and the exciton density) in two-photon spectroscopy. Beyond spectroscopy, the accurate separation of these processes directly impacts proposed quantum applications: while Rydberg exciton nonlinearities are envisioned for microwave–optical transducers and single-photon devices~\cite{walther2020controlling, assmann2020semiconductor, pritchett2024giant}, these require precise control over the real exciton population, as observed through SE. Thus, taking into account the SHG signal bias is an important step towards any future two-photon protocol. 


In this work, we carry out a systematic investigation of the light emitted under resonant two-photon excitation of $\mathrm{Cu_2O}$ Rydberg excitons. Using time-resolved measurements with optimized temporal resolution across a broad range of principal quantum numbers $n$, temperatures, excitation powers, and local crystal qualities, we cleanly separate the coherent SHG signal from the secondary emission (SE). This allows us to quantify their respective dependencies on $n$ and on external conditions, to identify the mechanisms that control each channel, and to establish robust experimental criteria for distinguishing SE from SHG in Rydberg-exciton spectroscopy.

\section{Experiment}

The experimental arrangement closely follows our previous time-resolved study of Rydberg exciton dynamics in $\mathrm{Cu_2O}$~\cite{Chakrabarti2025}. A high-purity natural $\mathrm{Cu_2O}$ single crystal with optically flat [111]-oriented faces was mounted in a cold-finger cryostat and held at a base temperature of \SI{4}{\kelvin}. Resonant two-photon excitation was provided by circularly polarized picosecond laser pulses ($\sim\SI{2}{\pico\second}$) near \SI{1142}{\nano\meter}, tightly focused to a \SI{10}{\micro\meter}-diameter spot on the \SI{50}{\micro\meter}-thick crystal. The emitted \SI{571}{\nano\meter}, corresponding to the $n=4-9$ Rydberg excitons, light was collected in transmission geometry and sent to a spectrometer equipped with either a cooled CCD (for steady-state spectra) or a streak camera (for time-resolved detection). The sub-picosecond temporal resolution of the streak camera enables clear discrimination between the instantaneous SHG response and the delayed secondary emission (SE) associated with radiative decay of the real exciton population.

To characterize spatial variations of crystal quality, we additionally performed wide-field resonant absorption imaging of the yellow $P$-excitons across the sample following Ref.~\cite{morin2024large}. The resulting quality map (Fig.~S2 of the Supplemental Material~\cite{SM}) was used to select three representative regions for detailed time-resolved two-photon excitation (TR-TPE) measurements. This allowed us to correlate the behavior of SHG and SE with the underlying local crystalline quality.

\section{Results and Discussions}

\begin{figure}[t]
\begin{center}
  \includegraphics[width=1.0\linewidth]{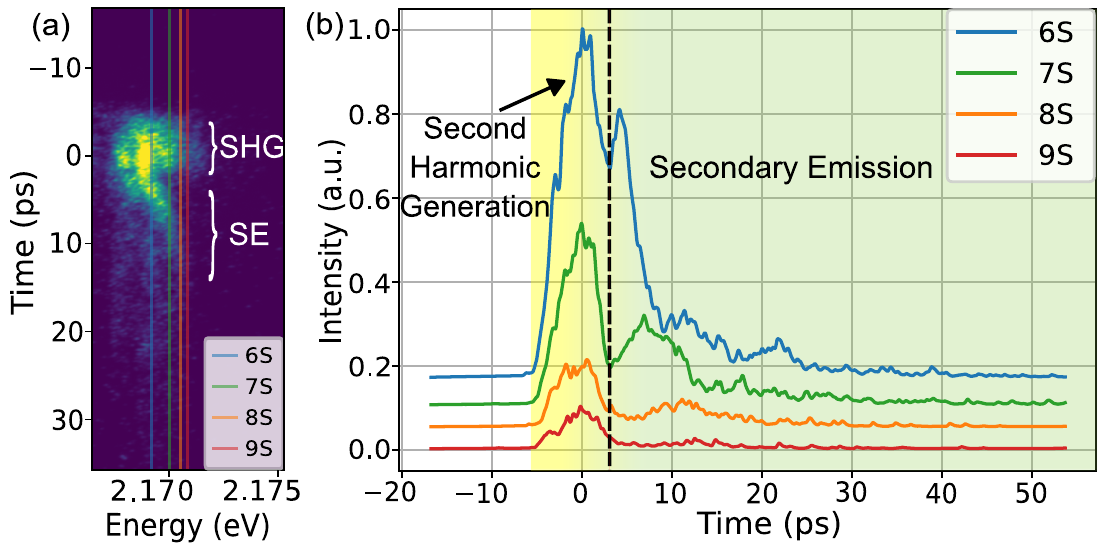}
  \caption{\textit{Time resolved emission}: \textit{(a)} Streak camera image of time (y-axis) and energy (x-axis) resolved signal from $7S$ and adjacent states. The average excitation power is 100~$\si{\milli\watt}$, centered on 1142.6~\si{\nano\meter} and inducing a signal around 2.1702~\si{\electronvolt}. The colored vertical lines indicate the energies of the selected $nS$ states. \textit{(b)} Time traces of the $nS$ states shown in (a), vertically shifted for clarity. The black dashed line indicates the transition from SHG (yellow) to SE (green).}
  \label{fig1}
  \vspace{-0.5cm}
\end{center}
\end{figure}

Time-resolved two-photon excitation (TR-TPE) allows us to disentangle the coherent $\chi^{(2)}$–driven SHG from the $\chi^{(3)}$ process that generates real excitons and their secondary emission (SE). As noted earlier, bulk $\chi^{(2)}$ vanishes in centrosymmetric $\mathrm{Cu_2O}$, so SHG is only weakly allowed along low-symmetry axes such as [111] or [112]~\cite{Mund2018high, Mund2019Second}. Under two-photon excitation tuned to the yellow Rydberg series, these higher-order SHG pathways are strongly enhanced, while the same photons also excite even-parity S and D excitons that subsequently relax and emit as SE.

Figure~\ref{fig1}(a) shows a representative time–energy streak image of the emitted light, where these two channels coexist and can be separated by their distinct timescales. The excitation was performed using an average excitation power of 100~$\si{\milli\watt}$ at half the energy of the $7S$ state. Owing to the finite spectral width of the pump pulse and the small separation between Rydberg states, many adjacent states are simultaneously excited. The energies of the excited $nS$ states are indicated by full lines on Fig.~\ref{fig1}(a) while their corresponding time traces are presented in Fig.~\ref{fig1}(b). The time traces $I(t)$ are obtained from the time-energy pictures by integrating the energy across the relevant segment for the state considered. In both panels, the SHG and SE are clearly separated along the time direction, especially for high $n$ due to their radically different dynamics. The SHG dynamics (yellow region) follows the expectations of nonlinear optics: it is synchronous with the laser pulse arrival, it lasts for exactly the pulse duration and follows its (symmetric) rise and fall time. As such, it is well described by a squared hyperbolic secant (or a Gaussian function, due to a small averaged jitter), and thus only exist at short times. On the other hand, the SE signal (green region) has a longer rise time (up to $\sim12~\si{\pico\second}$) and a much longer decay time (several tens of picoseconds). The slow decay corresponds to the Rydberg exciton lifetime and the oscillations visible in Fig.~\ref{fig1}(b) are coherent beating between the simultaneously excited states~\cite{Chakrabarti2025}. This dynamical distinction is quite evident for all exciton states except the lowest ($n<5$), for which the SHG and SE timescales are too similar to distinguish properly (see the Supplemental Material~\cite{SM} for details).

\begin{figure}[t]
\begin{center}
  \includegraphics[width=0.8\linewidth]{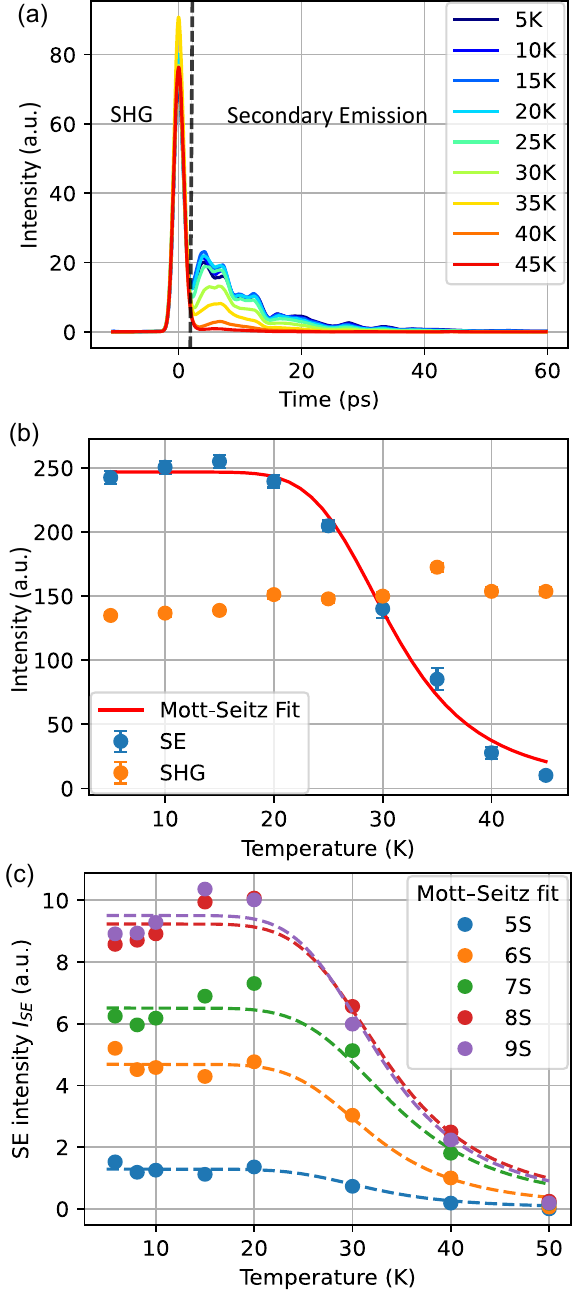}
  \caption{\textit{Temperature dependence}: \textit{(a)} Time-resolved emission traces recorded at different sample temperatures under resonant excitation centered on the $7S$ state. The average pump power was fixed at 100~$\si{\milli\watt}$. The black dashed line separates the SHG- (left) and SE- (right) dominated regions. \textit{(b)} Sample temperature dependence of the time-integrated SE (blue circles) and SHG (orange circles) contributions, numerically separated from the time traces. The red solid curve is a phenomenological Mott--Seitz-type fit to the SE intensity. \textit{(c)} Mott-Seitz fits for different $nS$ states at an average power of 50~$\si{\milli\watt}$. They all yield the same effective activation energies.}
  \label{fig2}
  \vspace{-0.5cm}
\end{center}
\end{figure}

\begin{figure*}[th]
\begin{center}
  \includegraphics[width=1\linewidth]{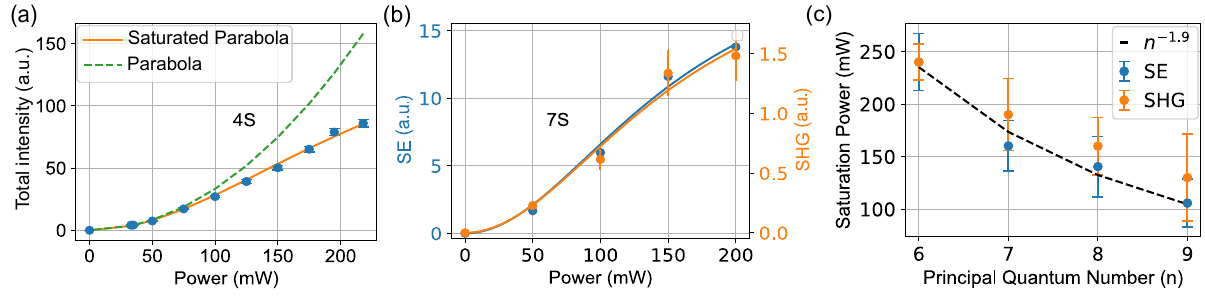}
  \caption{\textit{Power dependence.} \textit{(a)} Total integrated signal (SE+SHG) of the $4S$ state and \textit{(b)} separated SHG (orange) and SE (blue) intensities from the $7S$ state as a function of excitation power. Solid lines are quadratic saturation fits. \textit{(c)} Saturation powers versus the principal quantum number $n$. Orange and blue denote SHG and SE, respectively. The dashed line is a power law fit yielding $P_{\mathrm{sat}}\propto n^{-1.9}$.}
  \label{fig3}
  \vspace{-0.5cm}
\end{center}
\end{figure*}

As SHG and SE originate from fundamentally different physical processes, they are expected to respond differently to environmental changes. We first examine their temperature dependence. The sample temperature was increased from \SIrange[range-phrase = { to }, range-units = single]{5}{40}{\kelvin} in steps of \SI{5}{\kelvin}, and for each temperature a time trace of the emitted intensity was recorded. Figure~\ref{fig2}(a) shows a representative dataset for excitation tuned near the $7S$ resonance at an average power of \SI{100}{\milli\watt} (see Fig.~S3 in the Supplemental Material~\cite{SM} for additional examples). While the SE signal clearly decreases with increasing temperature, the SHG contribution changes only weakly.

The SHG component is therefore fitted at each temperature with a Gaussian centered at $t=0$ and a fixed temporal width equal to the pump pulse duration, leaving only the amplitude as a free parameter. Subtracting this contribution from the full time trace isolates the SE dynamics. The time-integrated intensities,
\mbox{$I_X=\int_{-\infty}^{+\infty} I_X(t)\,\mathrm{d}t$} with $X=\mathrm{SE}$ or $\mathrm{SHG}$,
are then obtained numerically. As summarized in Fig.~\ref{fig2}(b), the SHG intensity (orange circles) remains nearly constant over the investigated temperature range, whereas the SE intensity (blue circles) exhibits a pronounced reduction above about \SI{20}{\kelvin}, reaching roughly \SI{10}{\percent} of its low-temperature value at \SI{40}{\kelvin}. A similar loss of high-$n$ Rydberg-exciton visibility with increasing temperature has been reported in absorption spectroscopy, where it manifests as a reduction of spectral contrast and a decrease of the highest observable principal quantum number~\cite{kang2021temperature,Heckotter2025AdvQT}. We emphasize that, unlike absorption spectroscopy which probes the existence and spectral visibility of excitonic resonances, the SE intensity measured here reflects the radiative quantum yield of a real exciton population and is therefore sensitive to additional nonradiative loss channels.

We describe the SE quench using a phenomenological Mott--Seitz-type expression,
\begin{equation}
I_{\mathrm{SE}}(T)=\frac{I_0}{1+A\,\exp\!\left(-E_a/k_B T\right)},
\label{eq:MottSeitz}
\end{equation}
shown as the solid line in Fig.~\ref{fig2}(b). Fits performed at fixed excitation power yield an effective activation scale $E_a$ in the range $18$--$20~\si{\milli\electronvolt}$ for all investigated $nS$ states, as summarized in Fig.~\ref{fig2}(c). The absence of any systematic dependence of $E_a$ on $n$, despite the strong $(n-\delta)^{-2}$ scaling of the Rydberg binding energies, demonstrates that the SE quench is not governed by thermal ionization of a specific exciton state. Instead, the extracted activation scale reflects an energy that is essentially common to all states and must therefore be associated with phonon-assisted or defect-related processes. Its magnitude is comparable to characteristic optical-phonon energies in \(\mathrm{Cu_2O}\) (in particular, to the highest longitudinal-optical (LO) phonon, about \SI{19}{\milli\electronvolt}), suggesting that thermally activated phonon scattering can efficiently reduce the radiative quantum yield of the exciton population without necessarily destroying excitonic coherence. Possible mechanisms include phonon-assisted scattering into non-emissive momentum states outside the radiative light cone, or phonon-enabled access to nonradiative decay channels, such as ionization followed by recombination to the 1S state. In this context, it is worth noting that recent absorption studies emphasize that simple thermal dissociation by acoustic phonons cannot generally account for the temperature-limited visibility of Rydberg excitons at intermediate $n$, and that optical phonons and charged impurities play a dominant role~\cite{Heckotter2025AdvQT}.

In contrast to SE, SHG probes an early-time, coherence-driven polarization response and is therefore insensitive to late-time nonradiative decay channels. This explains why the SHG intensity remains almost temperature independent over the range considered here. At sufficiently high temperatures, beyond those explored in this work, SHG is also expected to decrease due to phonon-induced broadening and the eventual disappearance of the exciton resonance. Overall, these observations highlight that SE and SHG provide complementary information: SE probes the balance between radiative and nonradiative population dynamics, while SHG directly reflects the resonant nonlinear susceptibility of the Rydberg excitons.

\begin{figure*}[th]
\begin{center}
  \includegraphics[width=0.95\linewidth]{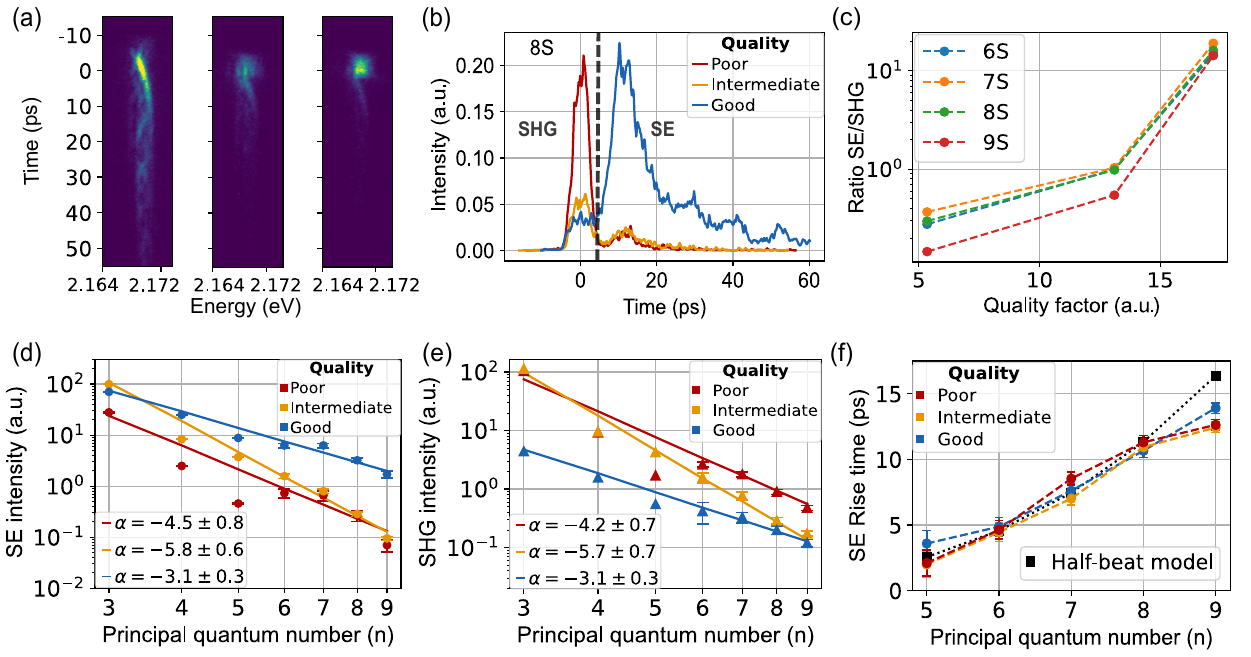}
  \caption{\textit{Dependence on sample quality}: \textit{(a)} Raw streak camera images obtained under the same conditions but varying sample quality  \textit{(b)} Typical time traces obtained from the streak images at different sample quality: good (blue), intermediate (orange) and poor (red). \textit{(c)} Ratio of SE to SHG intensities as a function of the crystal quality for different states. \textit{(d-e)} Scaling of the SE and the SHG intensity with $n$, for the three crystal quality. The pump power is fixed to 100~$\si{\milli\watt}$. Solid lines are power-law fits.  \textit{(f)} SHG-SE delay (SE rise time) versus $n$ for the three qualities. The black squares show a half-period for the beating between nS and nD states, versus $n$.}
  \label{fig4}
  \vspace{-0.5cm}
\end{center}
\end{figure*}

From here on, we focus on the base temperature of \SI{4}{\kelvin}. As both SHG and SE are initiated by a two-photon transition, they are expected to scale quadratically with the incident pump power at low intensity. This quadratic regime is clearly observed for all investigated states. At higher excitation powers, however, both signals deviate from the $P^2$ trend and progressively saturate beyond a characteristic power $P_{\mathrm{sat}}$, as visible in Figs.~\ref{fig3}(a-b). Using a saturated parabolic fit, we extract $P_{\mathrm{sat}}$ and plot it vs n in Fig.~\ref{fig3}(c). Remarkably, SHG and SE exhibit very similar saturation curves and nearly identical saturation powers, $P_{\mathrm{sat}}^{\mathrm{SHG}} \approx P_{\mathrm{sat}}^{\mathrm{SE}}$, for all accessible $n$. In a picture where Rydberg interactions block the excitation of closely spaced excitons, one would expect $P_{\mathrm{sat}}$ to follow the strong $n$-dependence of the blockade radius and scale approximately as $n^{-3.5}$, but the extracted $P_{\mathrm{sat}}(n)$ decreases more slowly (about $n^{-2}$). This suggests that additional nonlinear channels, such as the rapid formation of an electron–hole plasma from the high-energy tail of the pump spectrum or the generation of a blue–violet plasma by three-photon absorption~\cite{semkat2025interactionsS}, compete with or obscure a pure blockade behavior. A more quantitative identification of blockade physics would require a controlled separation of these contributions, which is beyond the scope of the present work. In any case, within our accessible range the SHG and SE signals display essentially the same power dependence, as expected from the fact that Rydberg interactions act instantaneously so long as the excitation density (real or virtual) is high enough. Therefore, the power scaling alone does not provide a reliable discriminator between the two processes.

We also examined the spatial and angular distributions of the emitted light using spatially and angle-resolved imaging (numerical aperture 0.34) in both transmission and reflection geometries. In the forward direction almost all of the signal is collected, with a spatial profile that closely follows the pump focus. The angular distributions of SHG and SE are very similar and are both more tightly confined around the optical axis (half-angle $\simeq\SI{3}{\degree}$) than expected from simple diffraction of the pump beam (half-angle $\simeq\SI{9}{\degree}$). This narrowing is consistent with the requirement that emission follows the [111] axis. In contrast, the backward-directed signal is approximately $250$ times weaker than the forward signal, which agrees with the level expected from reflection of the forward-propagating light once absorption in the crystal is taken into account. We therefore infer that essentially no light is emitted intrinsically in the backward direction: both SHG and SE are highly directional along the forward laser axis. This is consistent with recent observations of coherence in this system~\cite{Chakrabarti2025,farenbruch2025coherence}, for both early-time SHG and later-time SE, and indicates that the pump photon momentum is conserved in the emission process. Likewise, a related phenomenology is known from resonantly excited GaAs/AlGaAs quantum wells, where the resonant SE can include a coherent, phase-matched component~\cite{haacke1997resonant} (often discussed as resonant Rayleigh scattering from static disorder).

Another important factor influencing the signals is the static crystal environment. Strain and charged defects provide additional symmetry-breaking channels that can enhance the otherwise forbidden bulk $\chi^{(2)}$ response and thereby strengthen SHG~\cite{Mund2019Second}. At the same time, Rydberg excitons are extremely sensitive to local electric fields and disorder~\cite{heckotter2017scaling,kruger2020interaction}, so the SE signal is expected to be suppressed by charged impurities and associated field fluctuations. Thus, SHG and SE should display opposite trends with crystal quality. To test this, we recorded time-resolved signals in three regions of the crystal with different quality, labeled “good”, “intermediate”, and “poor” and indicated by blue, orange, and red markers on the quality map in Fig.~S2 of the Supplemental Material~\cite{SM}. The local quality parameter $Q(x,y)$ was obtained by energy-resolved imaging of the yellow $P$-excitons~\cite{morin2024large}, from which we estimate the corresponding charged-defect densities following Ref.~\cite{kruger2020interaction}: approximately $10^9$, $10^{10}$, and $10^{11}~\si{\centi\meter^{-3}}$ for the high-, medium-, and low-quality regions, respectively. The associated band-gap renormalization is negligible in this context (about \SI{0.4}{\milli\electronvolt} between the highest and lowest quality regions~\cite{kruger2020interaction}). Figure \ref{fig4}(a) displays representative streak images for the three regions  under identical excitation conditions at low temperature. Figure \ref{fig4}(b) shows the corresponding 8S time traces extracted from the streak images of each region. As anticipated, the early-time SHG peak increases markedly as the quality deteriorates, whereas the delayed SE tail is strongly suppressed. Coherent beating between neighboring Rydberg levels~\cite{Chakrabarti2025} is also rapidly washed out as the defect density increases. To highlight the contrasting behavior, Fig.~\ref{fig4}(c) shows the ratio of time-integrated SE and SHG intensities, $I_{\mathrm{SE}}/I_{\mathrm{SHG}}$, for several $nS$ states as a function of the local quality parameter $Q$. This ratio varies by nearly two orders of magnitude between the best and worst regions, but remains almost independent of $n$ for a given quality. Combined with the observation that SHG and SE share the same power dependence at low temperature, this demonstrates that $I_{\mathrm{SE}}/I_{\mathrm{SHG}}$ primarily reflects the local crystal quality. This emphasizes the well-known requirement of high purity for Rydberg exciton experiments, and also implies that nominally similar measurements on different samples or sample regions may in practice probe different mixtures of SHG and SE.

The individual SE and SHG intensities are plotted as a function of the principal quantum number $n$ in Figs.~\ref{fig4}(d) and \ref{fig4}(e), respectively. In each region the data are fitted with a power law of the form $I\propto (n-\delta)^{\alpha}$, with $\delta\simeq 0.5$ the quantum defect for $nS$ states~\cite{rogers2022high}. In the best-quality region, the SE intensity follows the expected oscillator-strength scaling with $\alpha\simeq -3$~\cite{kazimierczuk2014giant}. In the intermediate- and poor-quality regions, the SE scaling becomes significantly steeper ($\alpha\simeq -4$ to $-6$), consistent with previous observations of a quality-dependent crossover beyond a certain defect density~\cite{kruger2020interaction} and with the strong increase of electric-field sensitivity at high $n$~\cite{heckotter2017scaling}. This behavior is also reminiscent of recent purification experiments, where capture-related processes exhibit a pronounced $n$-dependence of $n^{6.5}$~\cite{bergen2023large}, underscoring the fragility of high-$n$ excitons to impurity-induced perturbations. Interestingly, the SHG intensity exhibits a very similar $n$-dependence within each quality region, differing mainly by an overall, quality-dependent amplitude. This is consistent with the fact that both SHG and SE are resonantly enhanced by the same excitonic states and share the same two-photon coupling matrix elements, so that their leading $n$-dependence is governed by a common oscillator-strength-like factor. The steeper scaling in lower-quality regions then reflects the increased impact of disorder and static fields on these resonant amplitudes, while the pronounced difference in absolute magnitude between SE and SHG encodes the additional sensitivity of SE to nonradiative population loss.

From the time traces, we extract the delay between the SHG peak and the first maximum of the SE signal, which we use as a measure of the SE rise time. Figure~\ref{fig4}(f) summarizes this delay for the same three quality regions as a function of $n$. We observe a systematic increase from about \SI{2}{\pico\second} for the $5S$ state up to roughly \SI{14}{\pico\second} for $9S$. These delays are significantly longer than the pump-pulse duration and cannot be explained by a single population dynamics. We previously showed that the dominant coherent oscillations arise from quantum beating between the $nS$ and $nD$ states within the same principal quantum number~\cite{Chakrabarti2025}. Using the energy splittings $\Delta E_{nS,nD} = E_{nD} - E_{nS}$, the characteristic timescale of this coherent evolution is given by the beat period $T_{SD}=h/\Delta E_{nS,nD}$. Remarkably, the experimentally extracted delays are quantitatively consistent with half of this period, as shown by the black line in Fig.~\ref{fig4}(f). The appearance of a half-period rather than a full-period delay indicates that the initial coherent superposition is not constructive at $t=0$. In the present broadband excitation, the $nS$ and $nD$ states are detuned by several linewidth and therefore acquire different excitation phases through the complex excitonic response. As a result, the $nS$–$nD$ interference could be close to destructive at $t=0$ and the first constructive rephasing occurs after about half a beat period, which can be much longer than the laser pulse duration. This is confirmed quantitatively by our multi-population numerical study (see Ref.~\cite{zielinska2025quantum} and Fig.~S4 of the Supplemental Material~\cite{SM}). Importantly, the SE rise time is essentially independent of the local crystal quality, in stark contrast to the strong quality dependence of the SE amplitude. This shows that, provided excitons can be formed, the sub-\SI{20}{\pico\second} dynamics are governed primarily by intrinsic coherent evolution within the Rydberg manifold rather than by defect-assisted scattering or relaxation~\cite{haacke1997resonant}. Defects and nonradiative channels therefore predominantly affect the emission yield and late-time decay, but not the early-time build-up of the SE signal. 

Finally, we investigated the polarization of the SHG and SE signals for a fixed, circular pump polarization. A polarizer was added before the measurement apparatus to record polarization-filtered time traces for circular (left and right) and linear ($0-180~\si{\degree}$) polarizations. In almost all cases, the SE and SHG parts have identical polarizations. This polarization matches very well previous studies~\cite{Mund2018high,Mund2019Second,farenbruch2020prb} (maximum signal in the circular polarization crossed from the pump) and we can therefore conclude that both contributions obey the same optical selection rules. 

\section{Conclusion}

We have used time-resolved two-photon excitation of Rydberg excitons in $\mathrm{Cu_2O}$ to separate two distinct emission channels: secondary emission (SE) from the radiative decay of real exciton populations, and second-harmonic generation (SHG) arising from a virtual excitonic $\chi^{(2)}$ response. SHG follows the pump pulse and provides a direct probe of the coherent excitonic susceptibility, whereas SE appears with a delayed rise and provides a direct window into the post-pump exciton dynamics.

The temperature dependence shows a clear contrast between the two channels: the SHG intensity remains nearly constant, while SE is strongly quenched above about \SI{20}{\kelvin}. The SE quench can be captured by a Mott–Seitz-type law with an effective activation scale $E_a\sim\SIrange{18}{20}{\milli\electronvolt}$ that is nearly independent of $n$, pointing to the activation of nonradiative channels. SHG, being a virtual process, is largely insensitive to these late-time nonradiative channels over the investigated range.

We also studied the power and quality dependence of both signals. SHG and SE scale quadratically with excitation power at low intensity and saturate at similar powers for all accessible $n$, so power scaling alone does not distinguish the two processes. In contrast, the ratio $I_{\mathrm{SE}}/I_{\mathrm{SHG}}$ varies by nearly two orders of magnitude between high- and low-quality regions while remaining essentially independent of $n$ and power. Finally, spatial and angular measurements show that both SHG and SE are highly forward-directed and obey similar polarization selection rules. 

Temporal discrimination is therefore essential whenever one aims to isolate SE and access the real exciton population dynamics. Together, these results provide a practical framework for separating and interpreting SHG and SE in $\mathrm{Cu_2O}$ Rydberg exciton experiments and offer guidelines for future studies targeting nonlinear and quantum-optical applications based on these states.

\begin{acknowledgments}
This work has been supported through the ANR grant ANR-21-CE47-0008 (PIONEEReX), through the EUR grant NanoX ANR-17-EURE-0009 in the framework of the "Programme des Investissements d’Avenir", through the Junior Professor Chair grant ANR-22-CPJ2-0092-01 and co-financed by the European Union through the European fund for regional development FEDER-FSE+ 2021- 2027 OCC008914. Support from National Science Centre, Poland (Project No. OPUS 2025/57/B/ST3/00334), is greatly acknowledged by SZ-R and DZ.
\end{acknowledgments}

\bibliographystyle{apsrev4-2}
\bibliography{biblio}

\end{document}


\title{Supplemental Material for ``Emission Dynamics of Rydberg Excitons in $\mathbf{\mathrm{Cu_2O}}$: Distinguishing Second Harmonic Generation from Secondary Emission''}

\author{Kerwan Morin}
\thanks{These authors contributed equally to this work.}
\affiliation{Univ Toulouse, INSA Toulouse, CNRS, LPCNO, Toulouse, France}
\author{Poulab Chakrabarti}
\thanks{These authors contributed equally to this work.}
\affiliation{Univ Toulouse, INSA Toulouse, CNRS, LPCNO, Toulouse, France}
\author{Delphine Lagarde}
\affiliation{Univ Toulouse, INSA Toulouse, CNRS, LPCNO, Toulouse, France}
\author{Maxime Mauguet}
\affiliation{Univ Toulouse, INSA Toulouse, CNRS, LPCNO, Toulouse, France}
\author{Sylwia Zielińska - Raczyńska}
\affiliation{Department of Physics, Bydgoszcz University of Science and Technology, Bydgoszcz, Poland}
\author{David Ziemkiewicz}
\affiliation{Department of Physics, Bydgoszcz University of Science and Technology, Bydgoszcz, Poland}
\author{Xavier Marie}
\affiliation{Univ Toulouse, INSA Toulouse, CNRS, LPCNO, Toulouse, France}
\author{Thomas Boulier}
\email{boulier@insa-toulouse.fr}
\affiliation{Univ Toulouse, INSA Toulouse, CNRS, LPCNO, Toulouse, France}

\date{\today}

\maketitle

\setcounter{figure}{0}
\setcounter{table}{0}
\setcounter{equation}{0}
\renewcommand{\thefigure}{S\arabic{figure}}
\renewcommand{\thetable}{S\arabic{table}}
\renewcommand{\theequation}{S\arabic{equation}}

\section{Data processing}
\label{sec:dataprocessing}

The raw data consist of time–energy streak images acquired with a spectrometer and streak camera. The horizontal axis encodes photon energy and the vertical axis encodes arrival time. For each dataset we first perform a standard preprocessing that includes:
(i) subtraction of a background image acquired with the pump blocked,
(ii) correction for the streak-camera and CCD dark counts,
and (iii) removal of slow drifts in the baseline by subtracting a low-order polynomial fitted to regions outside the excitonic features.

\subsection{Extraction of time traces for individual \texorpdfstring{$nS$}{nS} states}

To obtain the time traces $I_n(t)$ for individual $nS$ states, we integrate the streak image over a narrow spectral window centered on the corresponding exciton line. The window width is chosen such that it contains the full $nS$ absorption line while minimizing overlap with neighboring states; this is checked explicitly at each temperature and power setting. This integration yields a one-dimensional time trace
\begin{equation}
    I_n(t) = \int_{\Delta E_n} I(E,t)\, \mathrm{d}E,
\end{equation}
where $\Delta E_n$ denotes the energy window around the $nS$ resonance. The same procedure is applied for all temperatures, powers, and sample positions discussed in the main text.

\subsection{Separation of SHG and SE contributions}
\label{sec:SHGSEseparation}

Each time trace $I_n(t)$ contains an early-time peak due to second-harmonic generation (SHG) and a delayed tail due to secondary emission (SE). The SHG response follows the pump pulse and the overall instrument response, and is much shorter than the SE dynamics. We model the SHG contribution by a Gaussian centered at $t=0$,
\begin{equation}
    I_{\mathrm{SHG}}(t) = A_{\mathrm{SHG}} \exp\!\left[-\frac{t^2}{2\sigma_t^2}\right],
    \label{eq:SHGfit}
\end{equation}
with a fixed temporal width $\sigma_t$ equal to the independently measured pump-pulse width (convolved with the streak-camera response). The only free parameter is the amplitude $A_{\mathrm{SHG}}$, which is obtained from a least-squares fit of Eq.~(\ref{eq:SHGfit}) to the data in a narrow time window around $t=0$ (typically a few picoseconds where the SE contribution is still negligible).

The SHG-only contribution is then subtracted from the full trace,
\begin{equation}
    I_{\mathrm{SE}}(t) = I_n(t) - I_{\mathrm{SHG}}(t),
\end{equation}
to yield the SE dynamics. This procedure is robust as long as the SHG peak is temporally well-separated from the SE rise, which is the case for all states with $n\geq 5$ and for the experimental parameters used here.

The individual time-integrated contributions (SHG and SE intensities) are then obtained by numerical integration,
\begin{equation}
    I_X = \int_{t_{\mathrm{min}}}^{t_{\mathrm{max}}} I_X(t)\,\mathrm{d}t,
    \quad X = \mathrm{SE},\mathrm{SHG},
\end{equation}
where $t_{\mathrm{min}}$ and $t_{\mathrm{max}}$ define a fixed temporal window large enough to capture the full SE decay while avoiding regions contaminated by electronic artefacts at the beginning and end of the streak sweep. The same integration window is used for all datasets in a given series (e.g., for all temperatures or all powers), so that relative trends can be compared directly.

\subsection{Data processing for low-$n$ states ($n<5$)}

For the short-lived states $n=3$ and $4$, the SHG peak is not clearly separated in time from the SE contribution: the SE rise time is comparable to the instrument response. In this case a direct temporal decomposition into SHG and SE components is not reliable.

\begin{figure}[h!]
\begin{center}
  \includegraphics[width=0.75\linewidth]{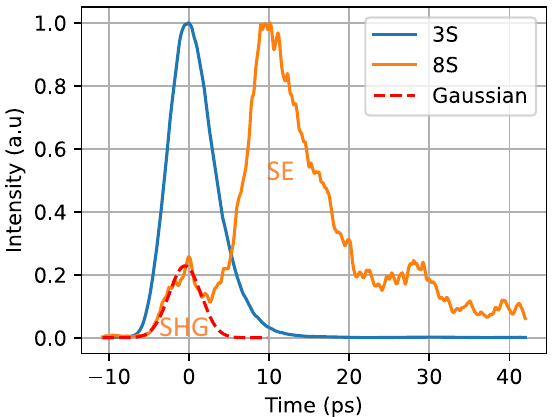}
  \caption{The time traces of 3S and 8S are shown in blue and orange, respectively. 8S exemplifies a case in which a temporal decomposition of SE and SHG can be performed. For 3S, an asymmetry is observed due to the SE lifetime. However, the decomposition of SE and SHG at an early time is unreliable as it falls within the instrument response time.   } 
  \label{TimeTraces}
\end{center}
\end{figure}

However, from the analysis of the higher states ($n\geq 5$) we find that at low temperature the ratio
\begin{equation}
    R = \frac{I_{\mathrm{SE}}}{I_{\mathrm{SHG}}}
\end{equation}
is essentially independent of $n$ within our experimental accuracy. We therefore infer the SE and SHG contents of the total $n=3$ and $4$ signals by assuming that the same ratio $R$ applies to these states as well. Writing the measured total intensity as
\begin{equation}
    I_{\mathrm{tot}} = I_{\mathrm{SE}} + I_{\mathrm{SHG}},
\end{equation}
and using $I_{\mathrm{SE}} = R\,I_{\mathrm{SHG}}$, we obtain
\begin{equation}
    I_{\mathrm{SHG}} = \frac{I_{\mathrm{tot}}}{1+R},
    \qquad
    I_{\mathrm{SE}} = \frac{R\,I_{\mathrm{tot}}}{1+R}.
\end{equation}
The value of $R$ used for this extrapolation is the average of the individually extracted ratios for $n=5$–$9$, and the spread between these values is propagated as an uncertainty on the inferred $n=3$–4 SE and SHG intensities. This procedure is only used for the low-temperature power-law scaling shown in Fig.~4(c–d) of the main text; all temperature-dependent and power-dependent analyses are restricted to states with directly separable SHG and SE contributions.

\section{Quality map and defect-density estimates}
\label{sec:qualitymap}

\begin{figure}[h!]
\begin{center}
  \includegraphics[width=1\linewidth]{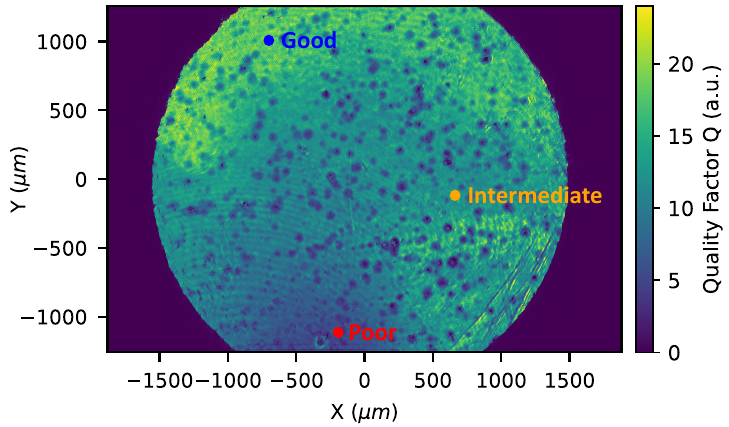}
  \caption{Quality map of the sample constructed from wide-field absorption spectra of the $8P$ state. Spatial coordinates are shown on the X and Y axes, with the color bar indicating the local quality factor $Q(x,y)$. The “good”, “intermediate”, and “poor” quality spots are indicated by blue, orange, and red circles, respectively.} 
  \label{Qmap}
\end{center}
\end{figure}

The spatial map of crystal quality shown in Fig.~\ref{Qmap} is obtained from wide-field resonant absorption imaging of the yellow $P$-excitons, following the method reported in Ref.~\cite{morin2024large}. Briefly, the sample is illuminated with a tunable, weak, quasi-collimated beam and the transmitted intensity is imaged onto a CCD camera. By scanning the photon energy across the $nP$ exciton resonances, we obtain a stack of transmission images $T(E,x,y)$ from which spatially resolved absorption spectra can be reconstructed.

For each pixel $(x,y)$ we automatically detect the peak profiles of the main yellow $P$-exciton lines and extract the peak optical density and linewidth. We then define a phenomenological quality factor $Q(x,y)$ that increases with increasing contrast and decreasing linewidth of the exciton resonance. In practice, we use
\begin{equation}
    Q(x,y) \propto \frac{A_P(x,y)}{\Gamma_P(x,y)},
\end{equation}
where $A_P$ and $\Gamma_P$ are the fitted amplitude and full width at half maximum of the $P$-exciton absorption.

Following the analysis of Krüger \emph{et al.}~\cite{kruger2020interaction}, we can relate the inhomogeneous broadening of the $P$-exciton line to the density of charged defects, which generate local electric fields and Stark-shift the exciton energies. Using the correlations reported there between linewidth and defect density, our measured linewidths correspond to average charged-defect densities on the order of $10^{9}$, $10^{10}$, and $10^{11}~\mathrm{cm^{-3}}$ in the high-, intermediate-, and low-quality regions, respectively. The associated band-gap renormalization between these regions is on the order of $0.4~\mathrm{meV}$, which is negligible on the scale of the laser linewidth and does not affect our resonance conditions. These numbers are intended as order-of-magnitude estimates; the main role of $Q(x,y)$ in this work is to provide a reproducible way to select and compare regions with markedly different levels of disorder.

\section{Low-power temperature study and activation-energy fits}
\label{sec:lowpowerT}

\begin{figure}[h!]
\begin{center}
  \includegraphics[width=1.0\linewidth]{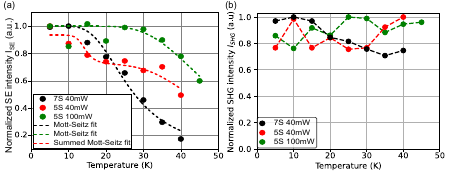}
  \caption{\textit{(a)} Temperature dependence of the time integrated SE contributions for $7S$ state at an average laser power of 40~mW (black circles), and the $5S$ state at excitation powers of 40~mW (red circles) and 100~mW (green circles). Dashed curves show Motz--seitz fits to the $7S$ 40~mW and $5S$ 100~mW  yielding activation energies of approximately $\sim10~\mathrm{meV}$ and $\sim20~\mathrm{meV}$, respectively. Red dashed line represents fit of $5S$ 40~mW data-points with sum of two Mott--Seitz-like contributions with fixed activation energies $E_{a1}=13.6~\mathrm{meV}$ and $E_{a2}=19.1~\mathrm{meV}$. \textit{(b)} Temperature dependence of the time integrated SHG contributions for $7S$ state at an average laser power of 40~mW (black circles), and the $5S$ state at excitation powers of 40~mW (red circles) and 100~mW (green circles).} 
  \label{temp}
\end{center}
\end{figure}

The main text discusses the temperature dependence of the SE and SHG intensities at an average pump power of 100-50~mW. The SE quench can be described by a Mott--Seitz-type law,
\begin{equation}
    I_{\mathrm{SE}}(T) = \frac{I_0}{1 + A\,\exp(-E_a/k_B T)},
    \label{eq:MSagain}
\end{equation}
with an effective activation scale $E_a$ of order $18$–$20~\mathrm{meV}$, essentially independent of the principal quantum number $n$ for $n\geq 5$.

To assess the robustness and physical meaning of $E_a$, we repeated the temperature-dependent measurements at several lower excitation powers (12.5~mW and 25~mW) for selected $nS$ states. This temperature dependence at lower power is shown in Fig.~\ref{temp}. For each power we follow exactly the same procedure as at 100~mW: streak images are processed as described in Sec.~\ref{sec:dataprocessing}, SHG and SE are separated in time, and the time-integrated SE and SHG intensity are extracted as a function of temperature between 5 and 40~K. The resulting $I_{\mathrm{SE}}(T)$ and $I_{\mathrm{SHG}}(T)$ curves are normalized to their low-temperature value for easier comparison.

We then fit each SE power-dependent dataset with Eq.~(\ref{eq:MSagain}), using $I_0$, $A$, and $E_a$ as free parameters. The fits describe the overall shape of the SE quench reasonably well for all powers, but the extracted activation energies show a marked and systematic dependence on power. As the average power is reduced, the apparent $E_a$ decreases and the onset of the SE quench shifts to lower temperatures. The spread in $E_a$ between the highest and lowest powers can reach a factor of two, whereas the statistical uncertainty of each individual fit is much smaller than this spread.

This power dependence demonstrates that $E_a$ should not be interpreted as a direct measurement of a single microscopic energy (such as the exciton binding energy of a specific state or a uniquely identified phonon). Instead, $E_a$ is an effective parameter that captures the combined influence of thermally activated processes (for example, LO-phonon-assisted scattering) and power-dependent channels (such as density-dependent nonradiative recombination, local heating, or defect neutralization). In particular, the observation that the SE becomes more robust against temperature at higher average power is qualitatively consistent with a reduction of defect-induced nonradiative channels at elevated Rydberg-exciton density, as discussed in connection with the ``purification'' mechanism in the main text.

For completeness, we have also tested alternative fitting forms, such as a sum of two Mott--Seitz-like contributions with fixed activation energies corresponding to the two dominant LO phonons in Cu$_2$O, and models where an additional power-dependent, temperature-independent nonradiative rate is added to the denominator. These more elaborate fits provide a description of similar quality and confirm that the simple single-activation form in Eq.~(\ref{eq:MSagain}) should be regarded as a compact empirical parameterization rather than a unique microscopic model.

\section{Theory support}
\subsection{Minimal model for the SE rise time}\label{minimalmodel}

The delayed maximum of the secondary emission (SE) relative to the second-harmonic generation (SHG) peak observed in Fig.~4(f) of the main text cannot be explained by a single-state population model. For an isolated exciton level driven by a short excitation pulse, the SE intensity would reach its maximum during or immediately after the pulse (up to a convolution with the pulse duration) and then decay monotonically. Such a picture predicts rise times comparable to the \SI{2}{\pico\second} pump duration and fails to account for the experimentally observed delays of up to \SI{15}{\pico\second}.

The key ingredient required to explain the data is the coherent excitation of multiple excitonic states by the broadband two-photon pulse. In particular, recent time-domain studies have shown that under resonant two-photon excitation the dominant coherent oscillations in \(\mathrm{Cu_2O}\) arise from quantum beating between the $nS$ and $nD$ states within the same principal quantum number manifold. Because the excitation pulse has a bandwidth comparable to or larger than the $nS$–$nD$ energy splitting for the states investigated here, it prepares a coherent superposition of these two states.

After the pulse, this superposition evolves freely, leading to oscillations at a frequency set by the energy splitting \(\Delta E_{nS,nD}\). The characteristic beat period is therefore
\begin{equation}
T_{SD} = \frac{h}{\Delta E_{nS,nD}} .
\end{equation}
The SE signal detected in a given spectral window can reach its first maximum after a finite delay if the initially prepared coherence is not maximally constructive at $t=0$. In the present broadband-excitation regime this is expected, since the $nD$ state is detuned from the laser center frequency and acquires a different excitation phase through the complex excitonic response. Residual chirp and phase structure of the excitation pulse further enhance these state-dependent phase differences. As a result, the initial $nS$–$nD$ interference can be close to destructive at $t=0$, and the first strong constructive rephasing occurs after approximately half a beat period,
\begin{equation}
t_{\max} \simeq \frac{h}{2\,\Delta E_{nS,nD}} .
\end{equation}

Using the measured $nS$–$nD$ energy splittings, this expression yields delays in the range \SIrange{3}{15}{\pico\second} for $n=5$–9, in good quantitative agreement with the experimentally extracted SE rise times. Because this timescale is set by the intrinsic Rydberg level structure, it is largely insensitive to excitation power and crystal quality, consistent with our observations. 

\subsection{Numerical model for the SE behavior}

\begin{figure}[h!]
\begin{center}
  \includegraphics[width=0.9\linewidth]{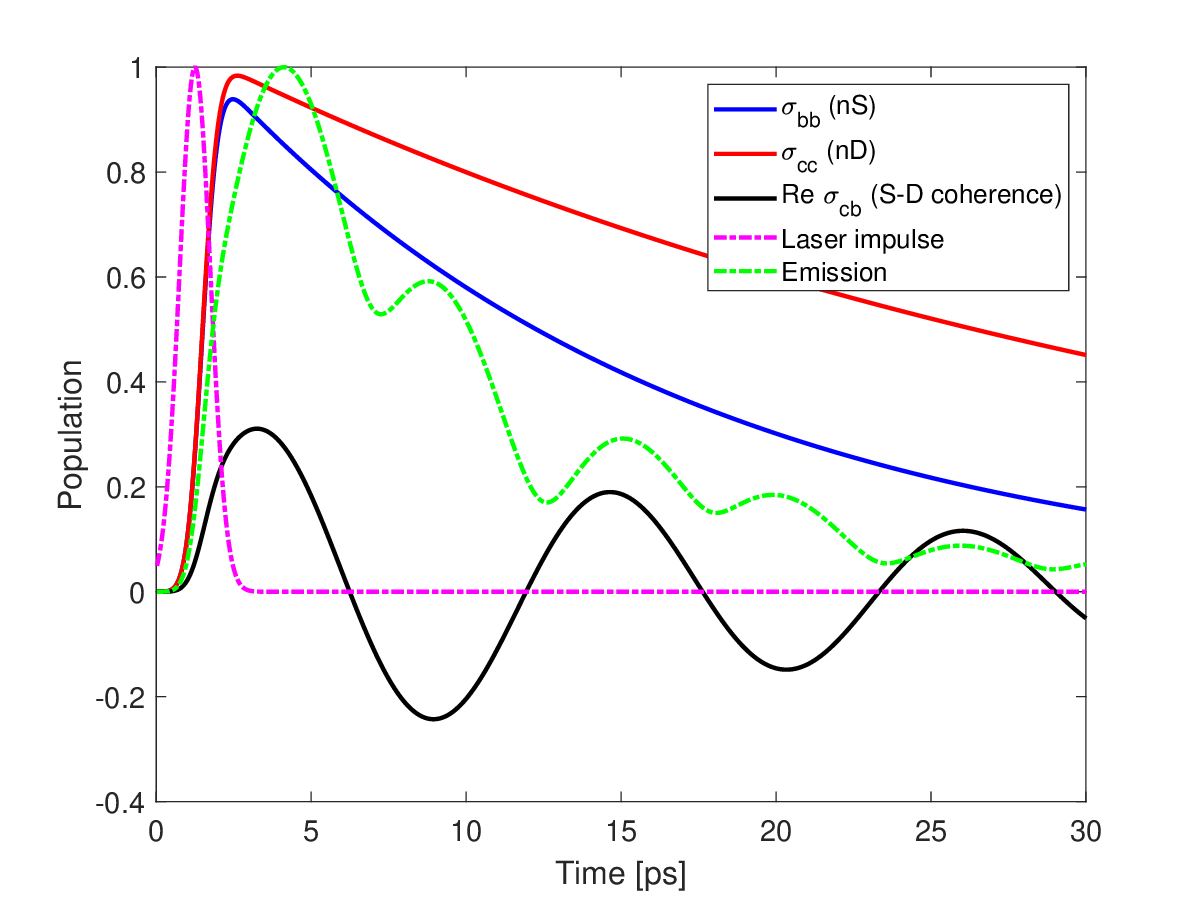}
  \caption{The time evolution of the real part of three density matrix elements, exciting laser intensity and emission intensity. All quantities normalized.} 
  \label{Theory_Beats}
\end{center}
\end{figure}

To describe the system dynamics, we use the density-matrix formalism $\rho(z,t)$, where $z$ denotes the position of an exciton. The time evolution is governed by the von Neumann equation with an additional phenomenological dissipation term $R$,
\begin{equation}\label{eq:vN}
i\hbar \dot{\rho}=[H,\rho]+R\rho.
\end{equation}
The Hamiltonian of the system, in the presence of excitation by an electromagnetic wave, is given by
\begin{eqnarray}
&&H=H_0+V+H_{vdW}=E_a|a\rangle \langle a|+\sum_j E_j|j \rangle \langle j|\nonumber\\
&&+\sum_j 2\Omega_j\hbar \cos(\omega t)\,[|a\rangle \langle j|+|j \rangle \langle a|],
\end{eqnarray}
where
\begin{equation}
\Omega_j=\frac{\varepsilon d_{aj}}{2\hbar}
\end{equation}
is the Rabi frequency for the $a \rightarrow j$ transition between the ground state and an excitonic state. Here, $E_j$ denotes the energies of the exciton states and $d_{aj}$ the corresponding transition dipole moments. In general, the Hamiltonian should also include the term $H_{vdW}$, which represents the van der Waals interaction arising from exciton--exciton interactions (Rydberg blockade). However, this interaction is not essential for the present discussion, particularly for the moderately excited Rydberg states considered in this work.

As a first approximation, we restrict our model to a three-level system consisting of the valence-band state $a$ and two upper excitonic states $b$ and $c$. We perform the rotating-wave approximation, i.e., we transform away the terms oscillating rapidly at the optical frequency: $\rho_{aj}=\sigma_{aj}\,e^{i\omega t}$, $\rho_{aa}=\sigma_{aa}$, and $\rho_{jj}=\sigma_{jj}$ with $j=b,c$. In the present case, owing to the symmetry properties of the valence and conduction bands of Cu$_2$O, two types of excitations are possible: dipole-allowed ($P$ excitons) and quadrupole-allowed but dipole-forbidden ($S$ or $D$ excitons). The one-photon (dipole-allowed) transitions are described by the coupling elements
\begin{eqnarray}
&&V_{aj}=-\varepsilon d_{aj},
\end{eqnarray}
whereas a dipole-forbidden excitation can be realized via a two-photon transition through a virtual, nonresonant state $v$,
\begin{equation}\label{eq:vab}
V_{aj}=\frac{W_{av}W_{vj}}{E_a-E_v-\tilde{\omega}}, \qquad \tilde{\omega}=2\omega,
\end{equation}
with $W_{av}=-\varepsilon d_{av}$. This effective second-order coupling given by Eq.~(\ref{eq:vab}) follows from the adiabatic elimination of terms involving the state $v$ from the complete set of density-matrix equations (see below), augmented by the virtual state $v$ and the corresponding additional couplings. If several nonresonant states $v_i$ contribute to the effective coupling, a sum over $v_i$ should be taken in Eq.~(\ref{eq:vab}).

Then, the von Neumann equation (\ref{eq:vN}), with  terms describing relaxations within the system, can be written as
\begin{eqnarray}\label{eq_glowne}
i\dot{\sigma}_{aa}&=&\Omega_{b}\sigma_{ba}+\Omega_{c}\sigma_{ca}-\Omega^*_{b}\sigma_{ab}-\Omega^*_{c}\sigma_{ac}\nonumber\\&+&i\Gamma_{ba}\sigma_{bb}+i\Gamma_{ca}\sigma_{cc},\nonumber\\
i\dot{\sigma}_{bb}&=&\Omega^*_{b}\sigma_{ab}-\Omega_{b}\sigma_{ba}-i\Gamma_{ba}\sigma_{bb}+i\Gamma_{cb}\sigma_{cc},\nonumber\\
i\dot{\sigma}_{cc}&=&\Omega^*_{c}\sigma_{ac}
-i\Gamma_{ca}\sigma_{cc}-i\Gamma_{cb}\sigma_{cc},\nonumber\\
i\dot{\sigma}_{ab}&=&\frac{1}{\hbar}(E_a+\hbar\omega-E_b)\sigma_{ab}+\Omega_{b}\sigma_{bb}+\Omega_{c}\sigma_{cb}\nonumber\\&-&\Omega_{b}\sigma_{aa}-i\gamma_{ab}\sigma_{ab},\nonumber\\
i\dot{\sigma}_{ac}&=&\frac{1}{\hbar}(E_a+\hbar\omega-E_c)\sigma_{ac}+\Omega_{c}\sigma_{cc}+\Omega_{b}\sigma_{bc}\nonumber\\&-&\Omega_{c}\sigma_{aa}-i\gamma_{ac}\sigma_{ac},\nonumber\\
i\dot{\sigma}_{bc}&=&\frac{1}{\hbar}(E_b-E_c)\sigma_{bc}+\Omega^*_{b}\sigma_{ac}-\Omega_{c}\sigma_{ba}-i\gamma_{bc}\sigma_{bc}.
\end{eqnarray}
The intensity of emission from the system is proportional to the mean  multipole (dipole $d_{ij}$ or quadrupole $q_{ij}$) moment $M$  between the ground and excited states, i.e. $I(t)\sim|\langle M \rangle|^2 $. It has the form
\begin{eqnarray}\label{exc_emis}
\langle M \rangle &=& Tr \rho M =(\sigma_{ab}M_{ba}+\sigma_{ac}M_{ca})e^{i\omega t}\nonumber\\
&+&(\sigma_{ba}M_{ab}+\sigma_{ca}M_{ac})e^{-i\omega t}\nonumber\\
&+&(\sigma_{bc}M_{cb}+\sigma_{cb}M_{bc}).
\end{eqnarray} 
Depending on whether the excitation is one- or two-photon, the emitted pulse originates from a dipole or quadrupole transition due to the selection rules.

An example result of the above calculation, performed for the $n=6$ exciton, is shown in Fig.~\ref{Theory_Beats}. We first focus on the elements $\sigma_{bb}$ and $\sigma_{cc}$ (blue and red curves), which represent the populations of the $S$ and $D$ exciton states, respectively. As expected, these populations increase during the excitation laser pulse, when energy is absorbed by the system, and subsequently decay exponentially according to the lifetimes of the involved states. The time evolution of the emission intensity (green curve) generally follows this behavior. However, there is a significant additional contribution from the off-diagonal element $\sigma_{cb}$ (black curve), which represents the coherence between the $nS$ and $nD$ exciton states and is responsible for the appearance of quantum beats. Initially, in Eq.~(\ref{eq_glowne}), we have $\sigma_{cb}=0$ and also $\dot{\sigma}_{cb}=0$. This implies that the quasi-sinusoidal quantum beat term, $\sim \sin(\omega_b t + \phi)$, must start with $\phi=0$ and then evolve with a frequency $\omega_b=\Delta E_{nS,nD}/\hbar$.

The position of the maximum in the emission intensity is determined by the interplay between coherent and incoherent contributions. Specifically, the incoherent emission reaches its maximum when the populations $\sigma_{bb}$ and $\sigma_{cc}$ are largest, which occurs near the end of the excitation pulse, when the energy absorption by the system is maximal. By contrast, the coherent contribution to the emission is strongest when the real part of $\sigma_{cb}$ is decreasing sharply. This takes place after approximately half of the beat period, as discussed in Sec.~\ref{minimalmodel}. This coherent contribution therefore dominates the temporal profile at early times, resulting in an emission delay that is close to half of the beat period and is largely independent of the excitation power.


\bibliographystyle{apsrev4-2}
\bibliography{biblio}